\documentclass[aps, prx, twocolumn, showpacs,superscriptaddress]{revtex4-1}
\usepackage{graphicx}
\usepackage{amsmath}
\usepackage{dsfont}
\usepackage{amssymb}
\usepackage{physics}
\usepackage{hyperref}
\usepackage{ulem}
\hypersetup{colorlinks=true,
	    final=true,
	    linkcolor=blue,
	    citecolor=blue,
	    filecolor=blue,
	    urlcolor=blue,
}

\newcommand{\LNO}[1]{La$_{3}$Ni$_{2}$O$_{#1}$}

\begin{document}

\title{Assessing the formation of spin and charge stripes in \texorpdfstring{La$_{3}$Ni$_{2}$O$_{7}$}{La3Ni2O7} from first-principles}
\author{Harrison LaBollita}
\email{hlabolli@asu.edu}
\affiliation{Department of Physics, Arizona State University, Tempe, AZ 85287, USA}
\author{Victor Pardo}
\affiliation{Departamento de Física Aplicada, Universidade de Santiago de Compostela, E-15782 Santiago de Compostela, Spain}
\affiliation{Instituto de Materiais iMATUS,
  Universidade de Santiago de Compostela, E-15782 Campus Sur s/n,
  Santiago de Compostela, Spain}
\author{Michael R. Norman}
\affiliation{  Materials Science Division, Argonne National Laboratory, Lemont, Illinois 60439, USA}
\author{Antia S. Botana}
\affiliation{Department of Physics, Arizona State University, Tempe, AZ 85287, USA}
\date{\today}

\begin{abstract}
We employ correlated density-functional theory methods (DFT + Hubbard $U$) to investigate the spin-density wave state of the bilayer Ruddlesden-Popper (RP) nickelate \LNO{7} which becomes superconducting under pressure. We predict that the ground state of this bilayer RP material is a single spin-charge stripe phase with in-plane up$^\prime$/up/down$^\prime$/down diagonal stripes with up$^\prime$/down$^\prime$ being low spin (formally Ni$^{3+}$: $d^7$) and up/down being high spin (formally Ni$^{2+}$: $d^8$). The main feature of this solution (that is insulating even at $U=0$) is the dominant role of $d_{x^{2}-y^{2}}$ bands around the Fermi level, which would become doped with the introduction of electrons via oxygen vacancies. In spite of the similarity with cuprates in terms of the dominant role of $d_{x^{2}-y^{2}}$ bands, some differences are apparent in the magnetic ground state of \LNO{7}: the antiferromagnetic out-of-plane coupling within the bilayer (linked to the $d_{z^2}$ orbitals forming a spin-singlet-like configuration) is found to be the dominant one while in-plane interactions are reduced due to the stripe order of the ground state. With pressure, this striped magnetic ground state remains similar in nature but the increase in bandwidth quickly transitions \LNO{7} into a metallic state with all the activity close to the Fermi level involving, to a large extent, $d_{x^2-y^2}$ orbitals. This is reminiscent of the cuprates and may provide key insights into how superconductivity arises in this material under pressure.

\end{abstract}

\maketitle

\textit{Introduction.--} The observation of superconductivity in low-valence layered nickelates, first in the infinite-layer compounds $R$NiO$_2$ ($R$ = rare-earth) \cite{Li2019superconductivity,Osada2020superconducting,Osada2021nickelate, Zeng2021superconductivity}, and more recently in the quintuple-layer material Nd$_6$Ni$_5$O$_{12}$ \cite{Pan2021superconductivity} has produced tremendous excitement in the condensed matter physics community in the past few years in the context of unconventional superconductivity. Structurally, these materials possess quasi-two-dimensional NiO$_2$ planes (analogous to the  CuO$_2$ planes of the cuprates) and belong to a larger family represented by the general chemical formula $R_{n+1}$Ni$_{n}$O$_{2n+2}$ where $n$ denotes the number of NiO$_2$ planes per formula unit along the $c$ axis. The discovery of superconductivity in this family of nickel oxide compounds completed a long endeavor to find cuprate analog materials in connection with high-T$_c$ superconductivity. Despite many structural, chemical, and electronic similarities to the cuprates \cite{Botana2020similarities}, the superconducting layered nickelates show some relevant differences, the most obvious one being their reduced superconducting critical temperatures T$_c$ $\sim$ 15 K \cite{Osada2020phase, Osada2020superconducting, Osada2021nickelate, Zeng2021superconductivity}, as well as their larger charge transfer energies (making them more ``Mott" like). Improved crystalline quality in samples of the infinite-layer nickelate \cite{Lee2022character}, as well as the application of hydrostatic pressure \cite{Wang2022Press} have shown modest increases in T$_{c}$, but they still remain far from typical cuprate values.

\begin{figure}
    \centering
    \includegraphics[width=\columnwidth]{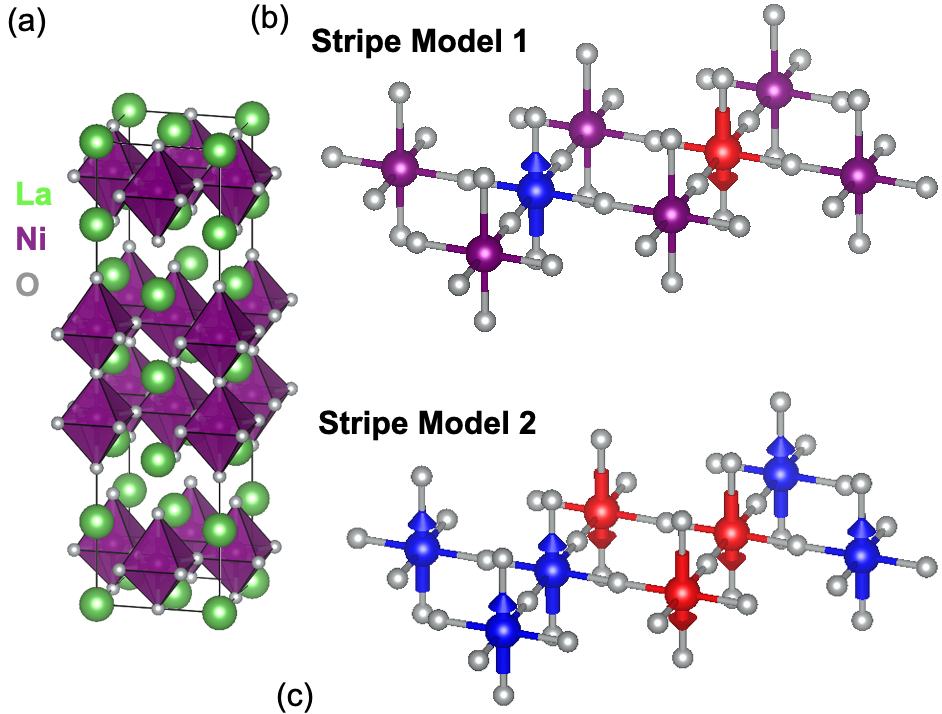}
    \caption{Crystal structure of \LNO{7} and the two more likely stripe models based on experimental data \cite{chen2024rixs,xie2024neutrons,chen2023musr,kakoi2024nmr}. (a) Crystal structure of bilayer \LNO{7} (shown in the unit cell of the \textit{Fmmm} space group used in our calculations) where La, Ni, and O atoms are denoted by green, purple, and grey spheres. (b) Stripe Model 1 consisting of a single spin-charge stripe (up/0/down/0) where ``0'' denotes spinless Ni sites (purple). (c) Stripe Model 2 consisting of a double spin stripe (up/up/down/down).}
    \label{fig:rixs-models}
\end{figure}

Recently, a breakthrough T$_{c}$ near 80 K  has been reported in the bilayer Ruddlesden-Popper (RP) nickelate \LNO{7} under pressure (P $\sim 14-42$ GPa) \cite{sun2023superconductivity,hou2023emergence,zhang2023exps}. The bilayer RP, \LNO{7},  differs from the previously reported superconducting nickelates in that it belongs to the parent $R_{n+1}$Ni$_{n}$O$_{3n+1}$ series \cite{Greeenblatt1997ruddlesden}. As such, it possesses layers  of NiO$_{6}$ octahedra (Fig.~\ref{fig:rixs-models}(a)) rather than having the square planar environment of the reduced RP phases. Moreover, the average oxidation state of the Ni ions is $2.5+$, corresponding to an average $3d^{7.5}$ filling, far from the $\sim3d^{8.8}$ filling where $T_c$ is optimal for the reduced RP phases. This has given rise to a plethora of experimental \cite{filamentary,chen2023musr,kakoi2024nmr, wang2023i4mmmexp} and theoretical work \cite{zhang2023electronic, chen2023critical, lechermann2023electronic, christiansson2023correlated, luo2023bilayer, gu2023effective, shen2023effective, zhang2023electronic, wú2023charge, yang2023possible, liu2023swave, zhang2023structural, qu2023bilayer, yang2023minimal, zhang2023trends, lu2023superconductivity, tian2023correlation,vortex2023huang, jiang2023screening, liao2023correlations, liao2023interlayer,oh2023tj,qin2023singlets, sakakibara2023hubbard, shilenko2023, sakakibara2023theoretical,yi2024antiferromagnetic}. In the theory context, special emphasis has been placed on the multi-orbital nature of the electronic structure, particularly on the pivotal role played by the $d_{3z^{2}-r^{2}}$ ($d_{z^2}$) orbitals and their strong antiferromagnetic (AF) coupling between layers.

Given that magnetism is a fundamental feature of the undoped cuprates, it is crucial to understand the nature of the magnetic ground state of \LNO{7} and its evolution under pressure. Resistivity measurements have shown a kink-like transition at $\sim$ 153 K, which responds to an external out-of-plane magnetic field, implying the existence of a spin-density wave (SDW) transition \cite{liu2023evidencesdw,wu2001magnetic}. Recent $\mu$SR  experiments suggested that a static long-range magnetic order emerges below 148 K, consistent with an SDW
internal field distribution \cite{chen2023musr, Khasanov2024musr}. Traces of a possible density wave have also been discovered by nuclear magnetic resonance (NMR) experiments \cite{kakoi2024nmr}. Furthermore, investigations utilizing resonant inelastic X-ray scattering (RIXS) and neutron scattering techniques on both single-crystal and powder samples of \LNO{7} \cite{xie2024neutrons, chen2024rixs} have corroborated the presence and characteristics of the spin-density wave (SDW) order below approximately 150 K. Based on this collection of different measurements, several magnetic structures that may underlie the SDW state of \LNO{7} have been proposed. Among the conjectured spin models, the most likely ones consist of a single spin-charge stripe (stripe 1), formed by an up/0/down/0 pattern of diagonal stripes, and a double spin stripe (stripe 2), consisting of an up/up/down/down pattern of diagonal stripes (see Figs.~\ref{fig:rixs-models}(b,c)). These stripe orders were proposed by RIXS measurements based on fits to the spin wave dispersion \cite{chen2024rixs} with both `stripe 1' and `stripe 2' qualitatively reproducing the experimental results. The `stripe 2' model was also proposed by an NMR study where no evidence for charge ordering was found \cite{kakoi2024nmr}. In contrast, $\mu$SR results support the single spin-charge stripe (`stripe 1'), excluding all possible magnetic structures without charge stripe formation \cite{chen2023musr, Khasanov2024musr}. Inelastic neutron scattering studies found that, within these two models, the `stripe 1' model was a better match to the observed excitations \cite{xie2024neutrons}. Hence, while all these experimental studies agree on the presence of spin-stripe formation and on the SDW propagation vector, $\mathbf{q}$ = (0.5, 0) (in orthorhombic notation), with AF coupling within the bilayer, there is no consensus on the specific type of stripe order in the plane.

Here, we use first-principles calculations to shed light on this issue by studying the energetics of the different spin-charge stripe models suggested in the literature to ultimately determine the magnetic ground state of \LNO{7} and its evolution with pressure. We find that the magnetic ground state at ambient pressure is a spin-charge stripe consisting of in-plane up$^\prime$/up/down$^\prime$/down diagonal stripes with alternating low spin (primed) and high spin (unprimed) Ni atoms (nominally Ni$^{3+}$ and Ni$^{2+}$, respectively). The favored out-of-plane magnetic coupling within the bilayer is found to be AF, as suggested by both neutron scattering and RIXS experiments \cite{xie2024neutrons,chen2024rixs}. This out-of-plane AF order is the dominant one and is linked to the strong overlap between the two $d_{z^2}$ orbitals, which leads to the formation of bonding-antibonding molecular orbitals. At ambient pressure, the magnetic ground state we find for \LNO{7} is insulating (consistent with the semiconducting-like behavior of transport data at low pressures \cite{taniguchi1995transport,zhang1994synthesis,kobayashi1996transport}) with the Ni-$d_{x^2-y^2}$ states and ligand O-$p$ orbitals dominating the low-energy physics with a high degree of hybridization between them. With pressure, the magnetic ground state remains similar in nature but the change in bandwidth quickly transitions \LNO{7} into a metallic state.  Much of the activity close to the Fermi level involves, to a large extent, the $d_{x^2-y^2}$ orbitals so their role in the electronic structure of this bilayer nickelate should not be disregarded.

\textit{Computational Methods.--} Our density-functional theory (DFT)-based calculations were performed using the all-electron, full-potential code {\sc wien2k} \cite{Blaha2020wien2k}. We used as a starting point the optimized structures obtained as described in Ref.~\onlinecite{labollita2023electronic} (with relaxed internal coordinates and experimental lattice constants) at both ambient pressure and 29.5 GPa. Given the large size of the supercells that would need to be implemented at ambient pressure in the experimentally reported \textit{Amam} structure, and to be able to perform a more direct comparison of pressure effects, we use the \textit{Fmmm} structures for both the ambient pressure case and the 29.5 GPa case.  In order to analyze the energetics of the two more likely stripe configurations described above (stripe 1 and stripe 2) using DFT calculations, an in-plane cell doubling was needed relative to the corresponding \textit{Fmmm} cells (the supercell space group is \textit{Pm}): the spin-charge stripe state (stripe 1) requires two types of Ni atoms to form magnetic/non-magnetic stripes. Overall, for our 2 $\times$ 1 cell, a hybrid between stripe 1 and stripe 2 (consisting of up/up/down/down diagonal stripes alternating high spin and low spin moments, i.e. up$^\prime$/up/down$^\prime$/down) turned out to be the most stable solution when adopting the Perdew-Burke-Ernzerhof (PBE) version of the generalized gradient approximation (GGA) as the exchange-correlation functional. Further structural relaxations (of internal coordinates only) were subsequently allowed in this magnetic ground state for this cell size within GGA-PBE. The resulting Ni-O distances (see Table \ref{distances}) show a modulation in the Ni-O bond length with longer (shorter) Ni-O distances for Ni atoms with a high (low) value of the spin moment.

\begin{table}
    \caption{Nearest neighbor Ni-O distances (in \AA) after relaxation for the ground state (C6) stripe solution at ambient pressure. There are four in-plane O neighbors for each Ni at the same distance. The two off-plane (apical) neighbors have different distances: a short bond between bilayers and a long bond pointing into the fluorite slab.}
    \begin{tabular*}{\columnwidth}{l@{\extracolsep{\fill}}cccc}
    \hline\hline
              &  Ni-O (planar)  & Ni-O (top apical) & Ni-O (bottom apical)\\
    \hline 
    Ni$^{2+}$ &   1.93          & 2.17              & 2.0 \\
    Ni$^{3+}$ &   1.90          & 2.23              & 1.99 \\
    \hline\hline    
    \end{tabular*}
    \label{distances}
\end{table}

With the optimized structures described above, we studied the trends in the energetics of different magnetic configurations with the on-site Coulomb repulsion ($U$) (including AF-A, AF-C, and a ($\pi$, 0)-stripe in addition to stripe 1 and stripe 2). Our DFT+$U$ calculations were performed with the local-density approximation (LDA) as the exchange-correlation functional and the around mean field (AMF) as the double counting correction term \cite{ldau_amf} based on the better description of experimental results for other nickelate compounds within this scheme \cite{Zhang2017large}.  We used a range of $U$ values applied to the Ni($3d$) orbitals of 1-5 eV, while the Hund's coupling ($J_{\mathrm{H}}$) was fixed to a typical value for transition-metal $3d$ electrons of 0.68 eV. A 10$\times$10$\times$10 $k$-point mesh was used for Brillouin zone integration. The basis set size was determined by $RK_{\mathrm{max}} = 7$ and muffin-tin radii (in atomic units) were set to 2.30, 1.86, and 1.65 for La, Ni, and O, respectively. Similar first-principles calculations as the ones presented here have been successfully used to predict and explain the charge-stripe and spin-stripe formation in the reduced trilayer nickelate La$_4$Ni$_3$O$_8$ \cite{botana2016charge,zhang2016stacked} that was subsequently verified by experiment \cite{zhang2019spin}.

\begin{figure*}
    \centering
    \includegraphics[width=1.75\columnwidth]{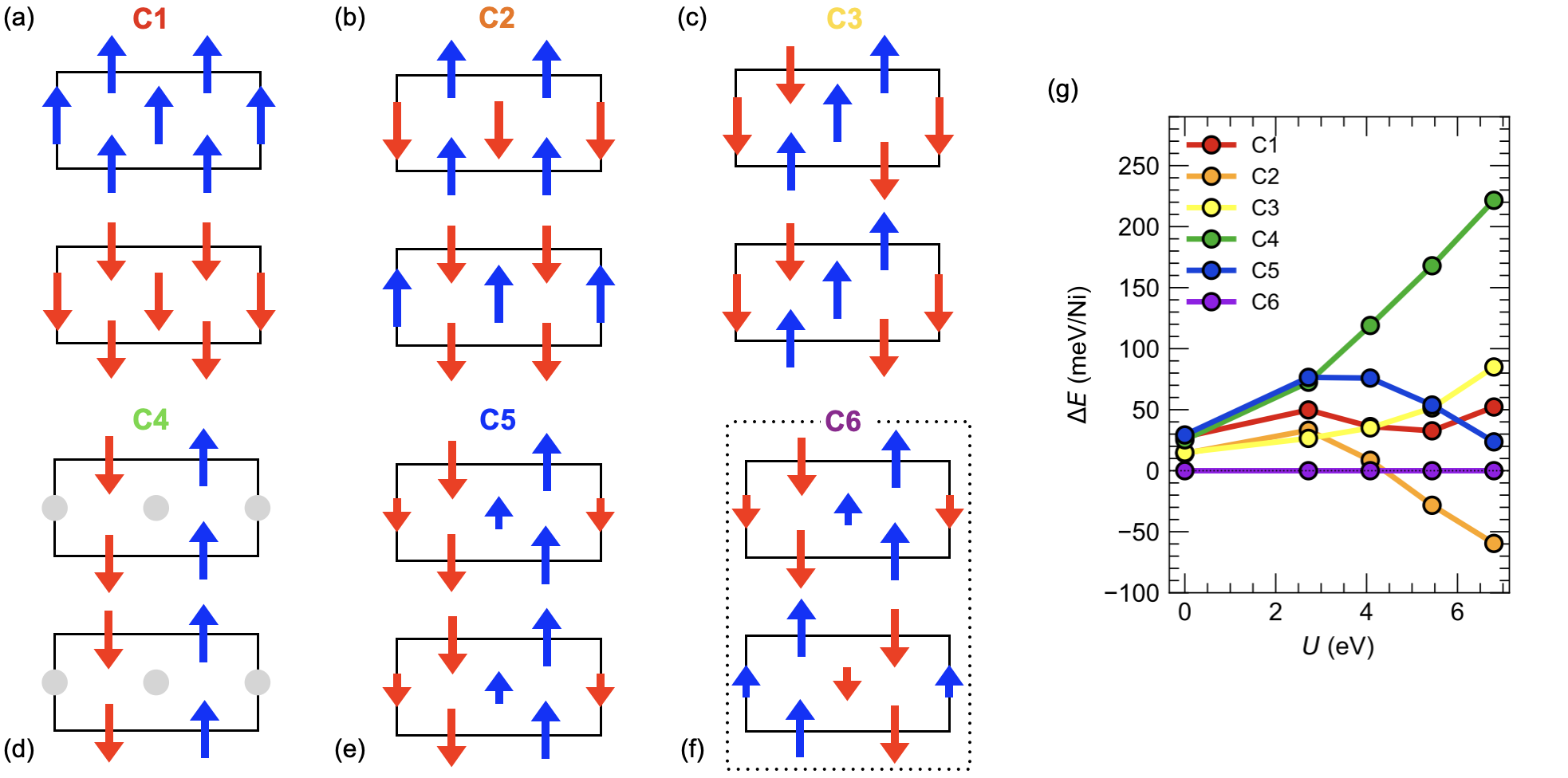}
    \caption{Stable (within DFT+$U$) spin configurations and corresponding energetics for \LNO{7} at ambient pressure. (a-f) Spin configurations represented in a magnetic supercell for the bilayer structural unit. (a) C1 corresponds to an A-type AF state, (b) C2 is a G-type AF state, (c) C3 corresponds to a ($\pi$, 0) stripe (up/down/up/down) with FM out-of-plane coupling, (d) C4 corresponds to the stripe model 1 (up/0/down/0), the ``0'' sites in gray being spinless (with FM coupling out-of-plane), (e) C5 corresponds to a hybrid between stripe 1 and stripe 2 models, consisting of diagonal up$^\prime$/up/down$^\prime$/down stripes alternating between high-spin (up/down, larger arrows) and low-spin moments (up$^\prime$/down$^\prime$, smaller arrows) also with FM out-of-plane coupling, (f) C6 is the same as C4 but with AF out-of-plane coupling. (g) Energetics computed within DFT+$U$ as a function of the Hubbard $U$ (the Hund's coupling $J_{\mathrm{H}}$ was fixed to 0.68 eV). The corresponding magnetic moments for all solutions are shown in Table \ref{tab:moments}.}
    \label{fig:p0-energetics}
\end{figure*}

\textit{Energetics of \LNO{7} at ambient pressure.--} We start by focusing on the energetics of the different magnetic configurations at ambient pressure we have converged which are shown in Figs.~\ref{fig:p0-energetics}(a-f). The configurations are as follows: (C1) corresponding to A-type AF order (consisting of ferromagnetic (FM) planes coupled AF out-of-plane), (C2) G-type AF order (consisting of AF checkerboard planes with AF coupling out-of-plane), (C3) a ($\pi$, 0) striped state, suggested in previous work as the magnetic ground state for La$_3$Ni$_2$O$_7$ \cite{zhang2023structural},
(C4) an up/0/down/0 pattern of diagonal stripes with FM coupling out-of-plane (identical to the stripe 1 model of Fig.~\ref{fig:rixs-models}) with 0 meaning non-magnetic (corresponding to a Ni$^{3+}$: $d^7$ ion) and up/down corresponding to a Ni$^{2+}$: $d^8$ ion, (C5) an up/up/down/down pattern of diagonal stripes (a hybrid between the stripe 1 and stripe 2 models of Fig.~\ref{fig:rixs-models}) with a low/high spin moment pattern with FM out-of-plane coupling (we denote this as up$^\prime$/up/down$^\prime$/down, primed corresponding to low-spin Ni$^{3+}$, and unprimed to high-spin Ni$^{2+})$, (C6) equivalent to C5 but with AF coupling out-of-plane. We summarize a few observations before carefully describing the magnetic ground state. Overall, the stripe 2 state suggested in Ref.~\onlinecite{chen2024rixs} can only be converged with moment disproportionation (as we describe in detail below) and the stripe 1 model can only be converged when the out-of-plane coupling is FM (when the out-of-plane coupling is set to be AF, the moments for the nominally non-magnetic Ni ions become non-zero and the solution converges instead to C6).  Importantly, our calculations indicate that the ground state for $U\leq$ 4 eV is the C6 stripe configuration (as reflected in the energy differences shown in Fig.~\ref{fig:p0-energetics}(g)). This magnetic ground state consists of in-plane up$^\prime$/up/down$^\prime$/down diagonal stripes with alternating high spin/low spin Ni atoms and AF out-of-plane coupling. Note that the $U$ value from cRPA~\cite{christiansson2023correlated} calculations in \LNO{7} is $\sim$3 eV, well within the range of stability of the ground state spin-charge stripe phase we propose.  At higher values of $U$, the G-type order becomes more stable. Hence, our calculations also show that an AF out-of-plane coupling is energetically favored for all values of $U$, in agreement with neutron scattering and RIXS data \cite{xie2024neutrons}.

\begin{table*}
    \centering
    \begin{tabular*}{2\columnwidth}{c@{\extracolsep{\fill}}ccccccc}
    \hline
    \hline
        Hubbard $U$ (eV)  &           & C1 (AF-A)   &  C2 (AF-G)    &  C3 ($\pi$, 0)     &  C4    &   C5     &  C6  \\
        \hline
        0                 & Ni$^{2+}$ &      0.47       &   0.52        &     0.60           &  0.81  &   0.84   &  0.90 \\
                          & Ni$^{3+}$ &    0.47         &   0.52        &     0.60           &  0.04  &   0.00   &  0.62 \\
                          \hline
        2.7               & Ni$^{2+}$ &       1.08      &   1.05        &     0.97           &  1.17  &   1.10   &  1.15 \\
                          & Ni$^{3+}$ &  1.08           &   0.30        &     0.97           &  0.06  &   0.76   &  0.72 \\
                          \hline
        4.0               & Ni$^{2+}$ &        1.19     &   1.22        &     1.08           &  1.34  &   1.30   &  1.30 \\
                          & Ni$^{3+}$ &  1.19           &   0.22        &     1.08           &  0.06  &   0.77   &  0.69 \\
                          \hline
        5.4               & Ni$^{2+}$ &         1.25    &   1.26        &     1.16           &  1.43  &   1.31   &  1.38 \\
                          & Ni$^{3+}$ &  1.25           &   0.21        &     1.16           &  0     &   0.35      &  0.56 \\
                          \hline
        6.8               & Ni$^{2+}$ &         1.44    &   1.26        &     1.22           &  1.49  &   1.29   &  1.37 \\
                          & Ni$^{3+}$ &  1.44           &   0.17        &     1.22           &  0     &   0.17      &  0.36 \\
    \hline
    \hline
    \end{tabular*}
    \caption{Magnetic moments (in $\mu_B$) for the different magnetic configurations we have obtained for \LNO{7} at ambient pressure.}
    \label{tab:moments}
\end{table*}

\begin{figure*}
    \centering
    \includegraphics[width=2\columnwidth]{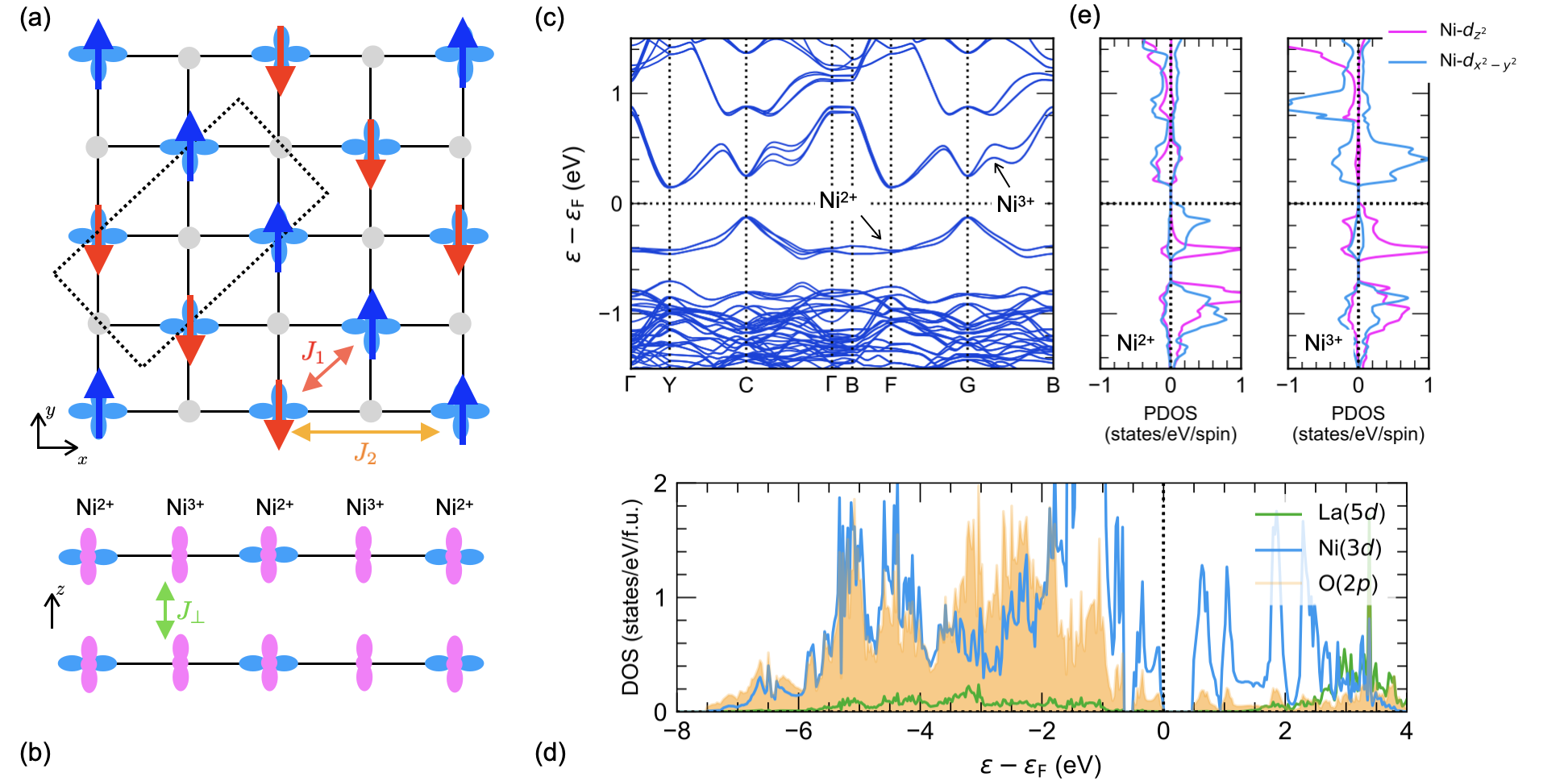}
    \caption{Electronic structure of the charge-spin stripe ground state of \LNO{7} at ambient pressure. Schematic representation of the stripe ground state in the (a) $ab$-plane with only the in-plane $d_{x^{2}-y^{2}}$ orbitals shown and (b) $ac$-plane, where the out-of-plane $d_{z^{2}}$ orbitals have been added. Note Ni$^{2+}$ sites correspond to $d_{z^{2}}^{1}$$d_{x^{2}-y^{2}}^{1}$ and Ni$^{3+}$ sites (gray) correspond to $d_{z^{2}}^{1}$$d_{x^{2}-y^{2}}^{0}$. (c) LDA+$U$ band structure in the magnetic ground state along high-symmetry lines ($U = 3$ eV, $J_{\mathrm{H}}=0.68$ eV) of the $Pm$ space group. High-symmetry points are: $\mathrm{Y}=(1/2, 0, 0)$, $\mathrm{C}=(1/2, 1/2, 0)$, $\mathrm{B}=(0,0,1/2)$, $\mathrm{F}=(1/2,0,1/2)$, and $\mathrm{G} = (1/2,1/2,1/2)$. (d) Corresponding atom-resolved density of states. (e) Orbital-resolved Ni-$e_{g}$ partial density of states for the nominally Ni$^{2+}$ (left) and Ni$^{3+}$ (right) sites. }
    \label{fig:ambientp_elecstr}
\end{figure*}

\textit{Electronic structure of the striped magnetic ground state of \LNO{7} and its evolution with pressure.--} We now analyze the electronic structure of the magnetic stripe state for $U\leq$ 4 eV (configuration C6 shown in Fig. \ref{fig:p0-energetics}). A summary of the electronic structure is shown in Fig.~\ref{fig:ambientp_elecstr} at $U= 3$ eV. The ground state charge-spin stripe configuration can effectively be described as an admixture of the stripe 1 and stripe 2 models proposed by RIXS (see Fig.~\ref{fig:rixs-models}(b,c)). The striped doubling of the unit cell gives rise to (effectively) two inequivalent Ni atoms with low and high spin moments in an up$^\prime$/up/down$^\prime$/down diagonal striped pattern with the closest ionic analog for these high/low spin Ni atoms corresponding to Ni$^{2+}$: $d^8$ ($S=1$) / Ni$^{3+}$: $d^7$ ($S=1/2$), respectively (see Fig.~ \ref{fig:ambientp_elecstr}(a,b)). The obtained magnetic moments (see Table \ref{tab:moments}) are consistent with the formal charges we have quoted, even though they are nominally reduced with respect to the ionic values due to hybridization with the ligands.  The charge/spin stripe ordered magnetic ground state we have obtained naturally opens a gap at the Fermi level (as clearly observed in the band structure and partial density of states (PDOS) in Fig.~\ref{fig:ambientp_elecstr}(c-d)). The gap opens up even at the GGA level (it simply increases in value when including a $U$). Such an insulating state is in agreement with the semiconducting-like behavior seen in transport data for La$_3$Ni$_2$O$_7$ samples close to stoichiometry \cite{sun2023superconductivity,hou2023emergence,zhang2023exps}.

We subsequently study the differences in orbital filling between the high-spin (nominally Ni$^{2+}$: $d^8$) and low-spin (nominally Ni$^{3+}$: $d^7$) focusing on the Ni-$e_{g}$ orbital-resolved PDOS (all the $t_{2g}$ states for both Ni ions are filled). The Ni$^{2+}$ ion has all of the $e_{g}$ majority spin states completely occupied (for both $d_{x^2-y^2}$ and $d_{z^2}$ orbitals) while the Ni$^{3+}$ only has the majority $d_{z^2}$ band occupied with the majority $d_{x^2-y^2}$ being completely empty. As can be seen in the partial density of states (PDOS), the occupied $d_{z^2}$ orbitals show a double-peak structure as a consequence of the bonding-antibonding splitting that arises due to quantum confinement in the structure that leads to the formation of these molecular orbitals \cite{Pardo2010quantum,pardo2011metal}. Importantly, the majority $d_{z^2}$ bands form a spin singlet out-of-plane: this many-body state is translated in DFT as a strong AF out-of-plane coupling. Indeed, the energy difference between the `singlet'-like (AF coupling out-of-plane) and `triplet'-like (FM coupling out-of-plane) configurations is large ($\sim$ 30 meV at the GGA level, rapidly increasing to $\sim$ 80 meV at $U= 3$ eV) as can be seen in Fig.~\ref{fig:p0-energetics}(g) by looking at the energy difference between the C5 and C6 configurations. This large energy difference agrees with the large $J_\perp$ quoted in Ref.~\onlinecite{chen2024rixs}, however we caution against considering this a superexchange interaction which would nominally decay with $U$.

Focusing on the $d_{x^2-y^2}$ orbitals, one ends up with a quarter-filled band (per Ni bilayer dimer) that alternates full/empty between neighboring Ni sites forming a charge stripe pattern in the plane (see Fig.~\ref{fig:ambientp_elecstr}(a,b)). Subsequently, the gap opens up between bands that are predominantly $d_{x^2-y^2}$ in character, with the conduction band nominally Ni$^{3+}$ and the valence band nominally Ni$^{2+}$ (as shown in Fig.~\ref{fig:ambientp_elecstr}(c)). Thus, the stripe pattern we have obtained gives rise to an electronic structure similar to that derived in the cuprates in terms of a half-filled $d_{x^2-y^2}$ orbital (such an orbital simply being quarter-filled per bilayer dimer in the \LNO{7} case instead, so the bilayer can be thought of as analogous to a single cuprate layer). Oxygen vacancies that naturally occur in La$_3$Ni$_2$O$_7$ would effectively electron-dope this material putting carriers in the $d_{x^2-y^2}$ orbitals above the Fermi level. 

Further, our calculations show that the favored in-plane coupling is AF, although in a more complex pattern than the standard checkerboard one obtains in the parent cuprates (in that context, we remark that G-type AF order is predicted to become stable at larger $U$). As shown in Fig.~\ref{fig:ambientp_elecstr}(a), there are two relevant in-plane magnetic exchanges: AF $J_1$ along the diagonal that corresponds to a direct ($\pi$) hopping between the half-filled $d_{x^2-y^2}$ bands of the Ni$^{2+}$ ions at a larger ($\times\sqrt{2}a$) distance when compared to cuprates, and AF $J_2$ that represents a superexchange via the low-spin Ni$^{3+}$ also at a larger  ($\times2a$) distance compared to cuprates. This should lead to reduced values of the estimated in-plane exchange constants with respect to standard cuprate values.

\begin{figure}
    \centering
    \includegraphics[width=\columnwidth]{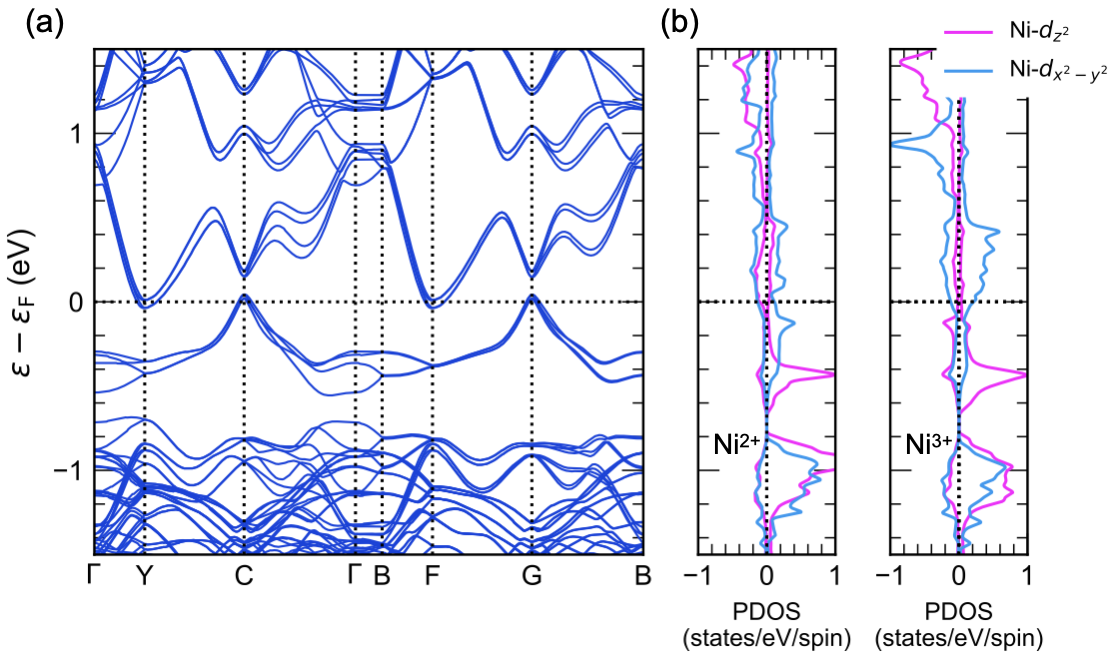}
    \caption{Electronic structure of \LNO{7} at P = 29.5 GPa in the proposed striped ground state within LDA+$U$ ($U = 3$ eV, $J_{\mathrm{H}} = 0.68$ eV). (a) Band structure along high-symmetry lines. (b) Orbital-resolved Ni-$e_{g}$ partial density of states for the nominally Ni$^{2+}$ (left) and Ni$^{3+}$ (right) sites.}
    \label{fig:29_5-ldauz}
\end{figure}

We now briefly analyze how the electronic structure of the stripe ground state we have obtained at ambient pressure (for $U\leq4$ eV) evolves under pressure, that is, when superconductivity arises in \LNO{7}. In the ambient pressure solution, both the valence and the conduction bands were mostly of $d_{x^2-y^2}$ parentage, with the valence (conduction) band coming mostly from nominal Ni$^{2+}$ (Ni$^{3+}$). With pressure, both these bands become broader and then overlap to form a metallic state with the Fermi level mostly composed of states of $d_{x^2-y^2}$ character as can be seen in Fig.~\ref{fig:29_5-ldauz}. The corresponding magnetic moments keep being consistent with the formal valences of a Ni$^{2+}$ and Ni$^{3+}$ ion quoted above (see Table \ref{tab:moments}). Such a tendency towards metallicity with pressure is obviously consistent with the appearance of superconductivity in \LNO{7} under pressure.

\textit{Nickelates vs. cuprates.--} We have obtained a striped magnetic ground state for \LNO{7} that involves alternating low-spin ($d^7$: $S=1/2$) and high-spin ($d^8$: $S=1$) Ni ions. An AF out-of-plane coupling is derived linked to the $d_{z^2}$ orbitals being strongly coupled between planes vis-à-vis molecular orbitals (this strong coupling effectively removes these orbitals from the low energy physics that is dominated by the $d_{x^2-y^2}$ states instead). Many of the characteristics of the spin-charge stripe state we have obtained for \LNO{7} at ambient pressure are typical of the cuprates where charge- and spin-density modulations also take place \cite{Keimer2015From}. Moreover, the occurrence of diagonal stripes is also reminiscent of the physics of the hole-doped single-layer nickelate La$_2$NiO$_4$ \cite{tranquada1996incommensurate,Wochner1998neutron,yoshizawa2000stripe}. It is important to emphasize that even though nominally La$_3$Ni$_2$O$_7$ corresponds to a quarter-filled $d_{x^2-y^2}$ band system (due to its average valence being Ni$^{2.5+}$), the ground state our calculations yield (consistent with RIXS, $\mu$-SR and neutron data) shows the picture is more complicated: the reorganization of $d_{x^2-y^2}$  states around the Fermi level coming from the two inequivalent Ni sites turns the electronic structure of this material into that of a (doped) half-filled $d_{x^2-y^2}$ band. In contrast to cuprates, this half-filled band nominally belongs to only one of the two Ni sites in the system, the other one providing the doping to it. Another important (and related) difference with the cuprates is that while cuprates have a large nearest-neighbor exchange and a much smaller next-nearest-neighbor exchange, in \LNO{7} the difference between these two interactions is much smaller. This leads to a more complex competition between magnetic phases in \LNO{7}. In addition, there is a dominant AF interlayer exchange in the nickelate
(linked to the $d_{z^2}$ orbitals forming a spin-singlet-like configuration), unlike cuprates. Overall, the striped ground state we have obtained in \LNO{7} renormalizes the bandwidths, the exchange interactions and many other characteristics when comparing it to cuprates, even though the low-energy physics seems to still be dominated by a single $d_{x^2-y^2}$ band around the Fermi level. 
Finally, we note that even though we have obtained within DFT+$U$ an up$^\prime$/up/down$^\prime$/down magnetic ground state where up/down corresponds to Ni$^{2+}$: $d^8$ ($S=1$) and up$^\prime$/down$^\prime$ to Ni$^{3+}$: $d^7$ ($S=1/2$), a description that corresponds to having a true $d_{z^2}$ out-of-plane singlet would be formed by $S=1/2$ ($d^8$) and $S=0$ ($d^7$) instead, akin to the so-called stripe-1 model in Fig.~\ref{fig:rixs-models}. The important point is that the magnetic ground state corresponds to alternating $d^8$-$d^7$ diagonal rows in both cases.

\textit{Summary and Conclusions.--} We have presented DFT+$U$ electronic structure calculations to understand the magnetic ground state of La$_3$Ni$_2$O$_7$ and its evolution with pressure. We find that the magnetic ground state at ambient pressure is a charge-spin stripe consisting of in-plane up$^\prime$/up/down$^\prime$/down diagonal stripes with primed corresponding to Ni$^{3+}$ ($d^7$) and unprimed corresponding to Ni$^{2+}$ ($d^8$)  ions. The energetics obtained from DFT shed light onto the different spin models proposed by RIXS, $\mu$-SR, and neutron scattering. This striped magnetic state is insulating at ambient pressure (but can easily become metallic via oxygen deficiencies) and, even though there is some relevant contribution of Ni-$d_{z^2}$ states, it is the  Ni-$d_{x^2-y^2}$ states that dominate the low-energy physics (on both sides of the Fermi level) and as such should not be disregarded in model calculations. Even though the stripe magnetic ground state has  important similarities with the cuprates, it also displays some important differences due to the complex in-plane striped magnetic configuration and the strong off-plane AF coupling.   With pressure, the striped magnetic ground state remains similar in nature but the increase in bandwidth quickly transitions \LNO{7} into a metallic state with all the activity close to the Fermi level involving, to a large extent, $d_{x^2-y^2}$ orbitals. This is reminiscent of the cuprates and may provide key insights into how superconductivity arises in this material under pressure.

\section*{acknowledgments}
 HL and ASB acknowledge the support from NSF Grant No.~DMR-2045826 and from the ASU Research Computing Center for HPC resources. VP acknowledges support from the Ministry of Science of Spain through the Project No.~PID2021-122609NB-C22 and the CESGA (Centro de Supercomputacion de Galicia) for the computing facilities provided. MRN was supported by the Materials Sciences and Engineering Division, Basic Energy Sciences, Office of Science, U.S.~Dept.~of Energy.

\bibliography{ref.bib}

\end{document}